\documentstyle[12pt,aasms4]{article}

\slugcomment{}

\lefthead{Halpern, Eracleous, & Forster}
\righthead{Optical & X-ray spec of 1E 0449.4-1823}

\begin{document}

\title{OPTICAL AND X-RAY SPECTROSCOPY OF 1E~0449.4--1823: \\
    DEMISE OF THE ORIGINAL TYPE~2 QSO}

\author{Jules P.\ Halpern\altaffilmark{1} and 
Michael Eracleous\altaffilmark{2,3}}
\affil {Department of Astronomy, University of California, Berkeley, CA 94720}
\affil {jules@astro.columbia.edu}

\and

\author{Karl Forster} \affil {Department of Astronomy, Columbia
University, 550 West 120th Street, New York, NY 10027}

\altaffiltext{1}{Permanent address: Department of Astronomy, Columbia
University, 550 West 120th Street, New York, NY 10027.}

\altaffiltext{2}{Visiting Astronomer, Cerro-Tololo International
Observatory, National Optical Astronomy Observatories, which is
operated by AURA, Inc., under a cooperative agreement with the
National Science Foundation.}

\altaffiltext{3}{Hubble Fellow.}

\received{9 September 1997}
\accepted{4 February 1998}

\bigskip
\centerline {To appear in the Astrophysical Journal}
\centerline {July 1, 1998, Vol. 501}

\begin {abstract}

New optical spectra of the original narrow-line quasar
1E~0449.4--1823\ show that it now has broad emission lines of
considerable strength, eliminating it as a ``type 2 QSO'' candidate.
Although broad emission-line components were probably present weakly
in 1981 and 1984, they have certainly increased in strength, and are
accompanied by Balmer continuum emission that makes the spectrum bluer
than it was previously.  We suggest that the behavior of
1E~0449.4--1823\ is the same as that of some Seyfert 1.8 and 1.9
galaxies, in which Goodrich attributed long-term variations of their
broad Balmer lines to dynamical motions of obscuring material located
in or around the broad-line region.  The optical continuum and broad
emission-line regions of 1E~0449.4--1823\ may still be partly covered
in our line of sight, which would explain its large forbidden-line
equivalent widths and flat $\alpha_{ox}$ relative to other
low-redshift QSOs.  Also present are apparent absorption features in
the broad Balmer lines and in Mg~II, which may be related to the past
obscuration and current emergence of the broad-line region.  However,
it is difficult to distinguish absorption from broad emission-line
peaks that are displaced in velocity; we consider the latter a
plausible competing interpretation of these peculiar line profiles.

An {\it ASCA\/}\ X-ray spectrum of 1E~0449.4--1823\ can be fitted with
a power-law of $\Gamma = 1.63^{+0.12}_{-0.09}$, intrinsic $N_{\rm H} <
9 \times 10^{20}$~cm$^{-2}$, and no Fe~K$\alpha$ line emission.  Its
2--10~keV luminosity is $6.7 \times 10^{44}$ ergs~s$^{-1}$.  Thus,
there is no evidence for Seyfert~2 properties in the X-ray emission
from 1E~0449.4--1823, which resembles that of an ordinary QSO.  With
regard to the still hypothetical type~2 QSOs, we argue that there is
little evidence for the existence of {\it any} among X-ray selected
samples.

\end{abstract}


\bigskip
\centerline {\bf 1. Introduction}

One basic fact about quasars is not yet understood, namely, why there
are virtually no narrow-line or ``type 2 QSOs,'' the high-luminosity
analogs of Seyfert 2 galaxies.  Every year or so, an X-ray discovered
object is advertised that might fit such a description, but its
qualifications are usually found to be lacking for one reason or
another.  A handful of such cases were described in recent papers by
Forster \& Halpern (1996) and Halpern \& Moran (1998), with generally
negative evaluations.  In this paper, we present new spectra of the
``original'' narrow-line QSO, the serendipitous X-ray source
1E~0449.4--1823\ at $z=0.338$ that was first to be described as such
(Stocke et al. 1982).  Although Stocke et al. referred to a possible
broad component of H$\beta$ in their discovery paper, and even
presented another spectrum (Stocke et al. 1983) showing a probable
weak broad Mg~II line in 1E~0449.4--1823\ shortly after their original
discovery, those observations have generally not been mentioned in
subsequent discussions of this object.  Instead, 1E~0449.4--1823\ is
invariably referred to as a Seyfert 2 galaxy or a narrow-line QSO
without qualification (e.g., Stephens 1989; Miller \& Goodrich 1990;
Keel et al. 1994; Elizalde \& Steiner 1995; Turner et al. 1997a, 1997b).

A weakness of the original studies of 1E~0449.4--1823\ was their lack
of coverage at H$\alpha$, which is sometimes the only broad emission
line that is clearly detectable in X-ray selected Seyfert galaxies
(e.g., Halpern, Helfand, \& Moran 1995).  Since this deficiency had
not yet been remedied to our knowledge, we undertook to obtain spectra
covering the Mg~II, H$\beta$, and H$\alpha$ regions of 1E~0449.4--1823\
in order to reassess its qualifications as a narrow-line QSO.  We also
reanalyzed an archival {\it ASCA\/}\ X-ray observation of
1E~0449.4--1823, previously published by Turner et al. (1997a, 1997b), in the
light of our new optical data.  The observations and results of this
investigation are reported below together with an interpretation of
the broad-line components of substantial strength that we discovered
in the optical spectrum.  Possible implications of the dearth of type
2 QSOs in general are also discussed.

\bigskip
\centerline {\bf 2. Optical Spectroscopy}

Optical spectra of 1E~0449.4--1823\ were obtained using the Kast
spectrograph (Miller \& Stone 1987) on the 3m Shane reflector of Lick
Observatory, and on the CTIO 1.5m telescope.  A log of the
observations is given in Table~1.  The spectrograph slit was oriented
at the parallactic angle in order to ensure spectrophotometric
accuracy.  A composite of the spectra is shown in Figure~1, after
standard reduction and dereddening by the extinction in this
direction, which is estimated to be $E(B-V) = 0.078$ from the neutral
hydrogen column density of Stark et al. (1992).  We measure a
heliocentric redshift of $0.3387 \pm 0.0001$ from the narrow emission
lines, which is consistent with the systemic galaxy redshift that can
be measured from the starlight that is definitely visible in the form
of the Ca~II~K~$\lambda$3933 absorption line.

\begin{figure}[h]
\plotfiddle{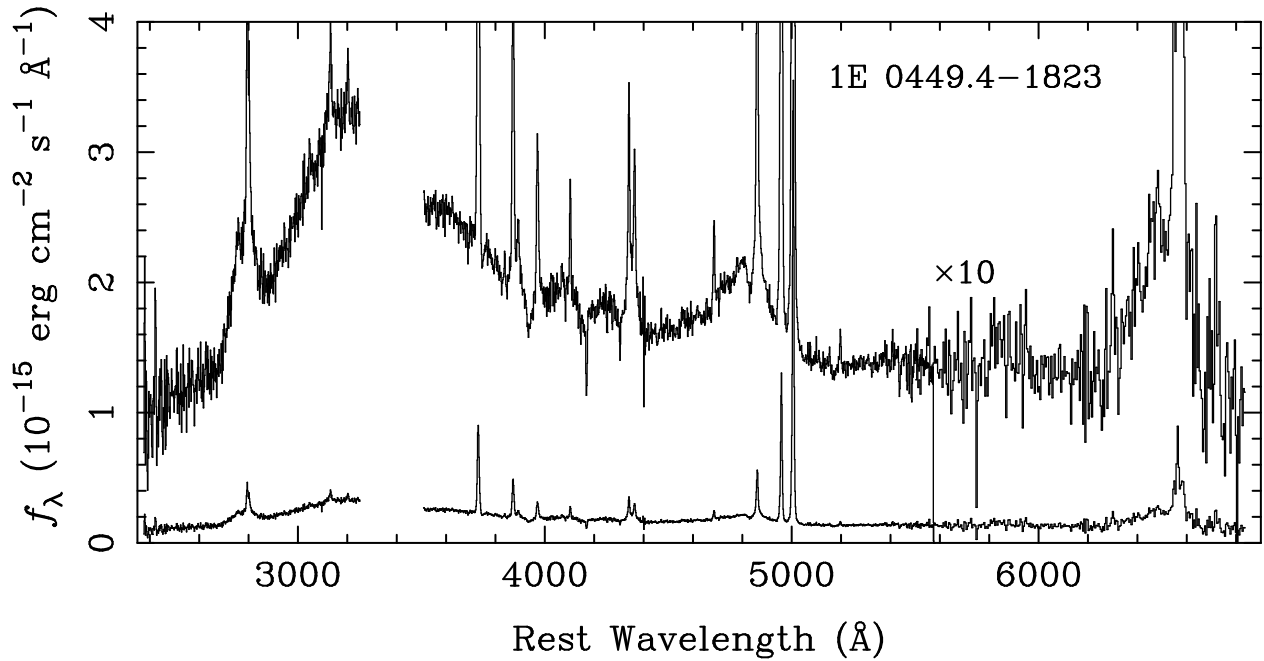}{3in}{270}{70}{70}{-270}{330}
\caption{Combined spectra of 1E~0449.4--1823\ from the
Lick 3~m ($< 5500$~\AA ) and CTIO 1.5m ($>5500$~\AA ).  The flux scale
refers to the lower trace.  The upper trace is the same data
multiplied by a factor of 10.  The gap in the spectrum is due to our
choice of dichroic filter for the Kast double
spectrograph.}
\end{figure}

\begin{table*}[b]
\begin{center}
TABLE 1

Log of Optical Spectroscopy of 1E~0449.4--1823

\begin{tabular}{clcccc}
\tableline
\tableline
Date & Telescope/Instrument & Exposure & Wavelength & Resolution & 
Slit width \\
(UT) &   & (s) &  (\AA )    & (\AA )    & ($^{\prime\prime}$) \\
\tableline
 1996 Oct 11 & Lick 3m/Kast Spec. & $2\times 2400$ & 3180--4526 & 4 & 2.0 \\
 1996 Oct 11 & Lick 3m/Kast Spec. & $2\times 2400$ & 4697--7468 & 5 & 2.0 \\
 1997 Jan 5  & CTIO 1.5m/RC Spec. & $3\times 1500$ & 7364--9152 & 3 & 1.8 \\
\tableline
\end{tabular}
\end{center}
\end{table*}

Immediately apparent in Figure~1 are broad emission lines of
substantial strength, including Mg~II, H$\beta$, and H$\alpha$.  The
broad H$\beta$ line, which here has rest EW = 55 \AA , was not obvious
in the spectra of Stocke et al. (1982) obtained in 1981, and that of
Stephens (1989) obtained in 1984.  However, the former authors noted
its possible presence, and it can be seen weakly upon close inspection
of the spectrum of the latter.  In addition to the improved
signal-to-noise ratio here, the broad lines have definitely increased
in strength since the early 1980s.  Another factor that hindered the
previous detection and measurement of the broad emission lines is
their large velocity width, $\approx 10,000$ km~s$^{-1}$ FWHM and
$\approx 20,000$ km~s$^{-1}$ FWZI.  In fact, this large velocity width
is a still a significant factor limiting the  accuracy to which the
Balmer lines can be measured; it is difficult to define a continuum
blueward of H$\beta$ because of blending with the higher-order Balmer
lines.  Accompanying those lines are an increase in the Balmer
continuum which is responsible for the broad bump in the near
ultraviolet.  This ``little blue bump'' was not present in the
previous spectra, leading Stocke et al. (1982) to describe
1E~0449.4--1823\ as a red object, with $U-B = -0.4$.  Although the
{\it rest frame} $U-B$ color is now --0.8 as estimated from our
spectrum, there is probably still significant reddening of the
continuum as indicated by its steepness in the neighborhood of the
Mg~II line.  In contrast, Grandi \& Phillips (1979) show spectra of QSOs
that are rising shortward of the Mg~II line.

Line intensity measurements for both broad and narrow components are
given in Table~2.  There is not much evidence for reddening in the
emission-line spectrum.  Both the narrow-line and broad-line Balmer
decrements are consistent with those of unreddened QSOs.  The broad
Mg~II/H$\beta$ ratio is 0.85, within the range 0.5--2.5 that is
ususally found in QSOs (Grandi \& Phillips 1979), and similar to those
of other X-ray selected AGNs (Puchnarewicz et al. 1997).

There is one noteworthy property of the spectrum of 1E~0449.4--1823.
A close examination of the broad emission lines shows that their
profiles (Figure~2) all have the same unusual structure, which can be
described either as an absorption feature blueshifted by
$-1900$~km~s$^{-1}$ from the galaxy rest frame, or else as a
well-defined broad emission-line peak blueshifted by
$-3500$~km~s$^{-1}$.  Although narrow absorption lines such as these
are common in the resonance lines of Seyfert galaxies, they are
virtually unknown in the Balmer lines, which makes us reluctant to
adopt the absorption-line hypothesis.  Alternatively, displaced broad
emission-line peaks are common in Balmer lines and Mg~II, but those
are usually found in radio galaxies or Seyferts of moderately high
radio luminosity (Eracleous \& Halpern 1994), whereas 1E~0449.4--1823\
is radio quiet, with flux densities of 1.1 mJy at 6~cm (Feigelson,
Maccacaro, \& Zamorani 1982) and 3.3 mJy at 20~cm (Condon et al. 1996,
the NRAO VLA Sky Survey).  [Although Ellsington, Yee, \& Green (1991)
refer to 1E~0449.4--1823\ as radio loud, this is clearly not the case,
as its monochromatic power at 20~cm is only $2.2 \times 10^{24}$
W~Hz$^{-1}$.  Throughout this paper we use $H_0 = 50$
km~s$^{-1}$~Mpc$^{-1}$ and $q_0 = 0$.]

One reason that it is difficult to decide between these two
descriptions of the spectra (absorption vs. displaced emission) is
that the contaminating narrow-line components are so strong.  Even
better spectra would be needed to distinguish between these
hypotheses.

\begin{table*}[h]
\begin{center}
TABLE 2

Emission-Line Measurements of 1E~0449.4--1823
\vskip 0.2cm
\begin{tabular}{lrrrr}
\tableline
\tableline
\quad Line Identification
 & Flux$^{\rm a}$ & Intensity$^{\rm b}$ & FWHM & EW$^{\rm c}$ \\
 & $F/F({\rm H}\beta)$ & $I/I({\rm H}\beta)$ & (km~s$^{-1}$ ) & (\AA ) \\
\tableline
Mg~II $\lambda$2795[narrow]          & 0.32 & 0.37 & ... & 6 \\
Mg~II $\lambda$2802[narrow]          & 0.17 & 0.20 & ... & 2 \\
Mg~II $\lambda$2798[broad]           & 1.75 & 2.04 & 10,200 & 47 \\
$ {\rm [O~III] } \lambda$3132              & 0.21 & 0.24 & ... & ... \\
He~II $\lambda$3203		     & 0.10 & 0.11 & ... & ... \\
$ {\rm [O~II] }  \lambda$3727                 & 1.61 & 1.73 & ... &  27 \\
$ {\rm [Ne~III] } \lambda$3869               & 0.72 & 0.77 & ... & ... \\
He~I $\lambda$3888, H$\zeta$	     & 0.29 & 0.31 & ... & ... \\
$ {\rm [Ne~III] } \lambda$3968, H$\epsilon$  & 0.46 & 0.48 & ... & ... \\
H$\delta$[{\it n}]                   & 0.35 & 0.36 & ... & ... \\
H$\gamma$[{\it n}]                   & 0.43 & 0.44 & ... & ... \\
H$\gamma$[{\it b}]                   & 0.86 & 0.89 & 9,200 & 21 \\
$ {\rm [O~III] } \lambda$4363                & 0.48 & 0.49 & ... & ... \\
\tableline
\end{tabular}
\end{center}
\end{table*}

\begin{table*}[p]
\begin{center}
TABLE 2 (cont.)
\vskip 0.2cm
\begin{tabular}{lrrrr}
\tableline
\tableline
\quad Line Identification
 & Flux$^{\rm a}$ & Intensity$^{\rm b}$ & FWHM & EW$^{\rm c}$ \\
 & $F/F({\rm H}\beta)$ & $I/I({\rm H}\beta)$ & (km~s$^{-1}$ ) & (\AA ) \\
\tableline
He~II $\lambda$4686                  & 0.15 & 0.15 & ... & ... \\
H$\beta$[{\it n}]                    & 1.00 & 1.00 & ... &  27 \\
H$\beta$[{\it b}]                    & 2.41 & 2.41 & 10,900 & 55 \\
$ {\rm [O~III] } \lambda$4959                & 2.43 & 2.42 & ... & 67 \\
$ {\rm [O~III] } \lambda$5007                & 7.22 & 7.17 & ... & 195 \\
$ {\rm [N~I] } \lambda$5199                  & 0.07 & 0.06 & ... & ... \\
$ {\rm [O~I] } \lambda$6300                  & 0.44 & 0.42 & ... & ... \\
$ {\rm [N~II] } \lambda$6548                 & 0.56 & 0.52 & ... & 19 \\
H$\alpha$[{\it n}]                   & 2.63 & 2.46 & ... &  87 \\
H$\alpha$[{\it b}]                  & 10.16 & 9.53 & 9,470 & 285 \\
$ {\rm [N~II] } \lambda$6583                 & 1.12 & 1.05 & ... &  37 \\
$ {\rm [S~II] } \lambda$6716                 & 0.54 & 0.50 & ... & ... \\
$ {\rm [S~II] } \lambda$6731                 & 0.24 & 0.22 & ... & ... \\
\tableline

\end{tabular}
\end{center}

\tablenotetext{a}{Observed flux relative to
$F({\rm H}\beta[n]) = 3.13 \times 10^{-15}$ ergs~cm$^{-2}$~s$^{-1}$.}

\tablenotetext{b}{Intensity corrected for Galactic $E(B-V) = 0.078$
mag, relative to \hskip 3cm $_{_{.}}$ $I({\rm H}\beta[n])~=~3.75~\times~10^{-15}$~ergs~cm$^{-2}$~s$^{-1}$.}

\tablenotetext{c}{Rest frame equivalent width.}

\end{table*}

\begin{figure}[p]
\plotfiddle{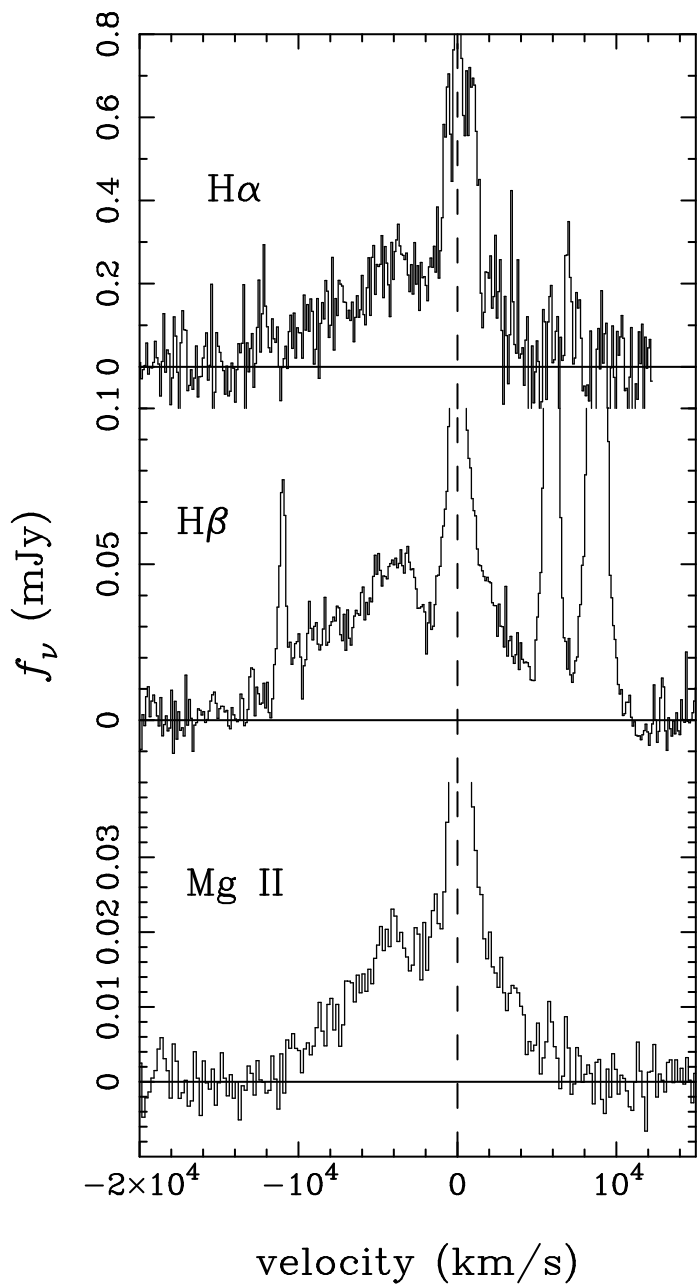}{6in}{0}{70}{70}{-220}{-50}
\caption{Continuum subtracted spectra of the broad
emission lines of 1E~0449.4--1823\ in velocity units.  Structure in
the broad emission lines can be described either as absorption
centered at $-1900$ km~s$^{-1}$, or a displaced broad emission-line
peak at $-3500$ km~s$^{-1}$.}
\end{figure}

\eject
\centerline {\bf 3. ASCA X-ray Observation}

1E~0449.4--1823\ was observed by the {\it ASCA\/}\ satellite on 1994
March 4--5.  Data obtained with the four instruments on board {\it
ASCA\/}\ were obtained from the archive, and were filtered following
the standard procedures described in {\it The ABC Guide to ASCA Data
Reduction\/}.  The SIS detectors were operated in 2-CCD mode, with the
target placed at the default 1-CCD position. Source counts in the SIS
images were extracted using a $3'$ radius circular region, and
background counts were collected from the entire chip, excluding a
$4'$ radius region centered on the target.  In the GIS images, source
counts were extracted using a region $6'$ in radius.  The GIS
background was measured in a source-free part of the image located the
same distance off-axis as 1E~0449.4--1823\ with an area equal to that
used to extract the target.  Useful exposure times and average
background-subtracted source count rates in the four {\it ASCA\/}\
detectors are listed in Table~3.

\begin{table*}[h]
\begin{center}
TABLE 3

{\it ASCA\/} Observation Summary

\begin{tabular}{lcc}
\tableline
\tableline
           & Exposure Time& Count Rate, 0.5--10 keV \\
 Instrument&           (s)&           (counts s$^{-1}$) \\
\tableline
SIS0 & 33,110 & $3.56 \times 10^{-2}$\\
SIS1 & 32,627 & $2.38 \times 10^{-2}$\\
GIS2 & 36,669 & $2.06 \times 10^{-2}$\\
GIS3 & 36,667 & $2.58 \times 10^{-2}$\\
\tableline
\end{tabular}
\end{center}
\end{table*}

For spectral fitting, the SIS and GIS spectra were rebinned to have at
least 20 counts (source plus background) per channel.  All four
detectors were modelled simultaneously, but for clarity of
presentation, the summed SIS and summed GIS spectra are shown in
Figure~3.  The spectra are modelled with a power law absorbed by a
column of neutral gas fixed at the Galactic value ($3.88 \times
10^{20}$ cm$^{-2}$; Stark et al. 1992).  A second absorption component
was included as a free parameter at the redshift of
1E~0449.4--1823. It was found that this additional intrinsic column
did not improve the fit and only an upper limit of $N_{\rm H}^{\rm
Int} < 8.8 \times 10^{20}$ cm$^{-2}$ (90\% confidence) can be derived
(see Figure~4).  The photon index is found to be $\Gamma = 1.63
^{+0.12}_{-0.09}$ and the observed flux between 0.5 and 10 keV is 
$1.6~\times~10^{-12}$~ergs~cm$^{-2}$~s$^{-1}$~keV$^{-1}$ (the average of
the four instruments).  The intrinsic luminosity of 1E~0449.4--1823\
in the rest frame 2--10~keV band is $6.7 \times 10^{44}$
ergs~s$^{-1}$.

\begin{figure}[h]
\plotfiddle{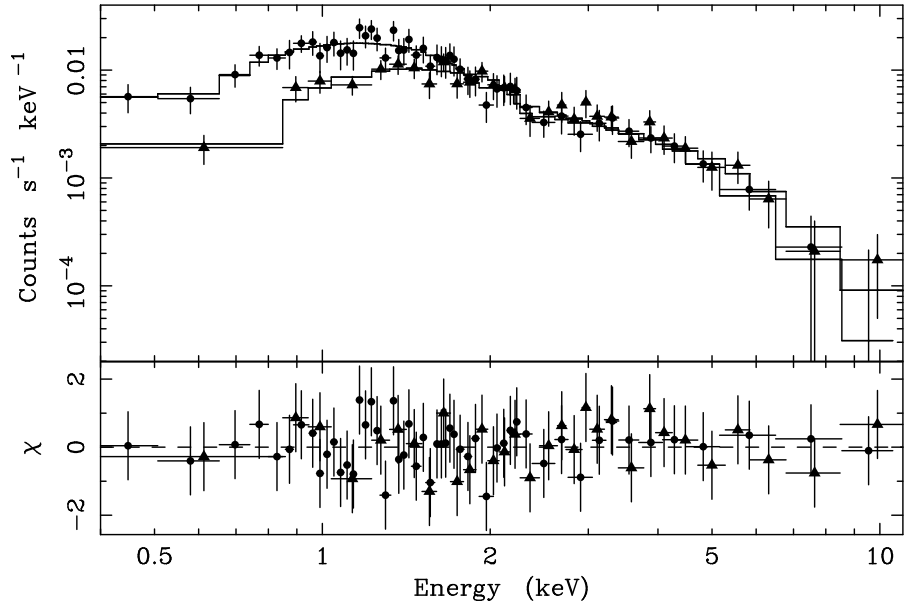}{3in}{270}{50}{50}{-180}{280}
\caption{The observed {\it ASCA\/}\ SIS (filled circles)
and GIS (filled triangles) spectra of 1E~0449.4--1823\ and best
fitting power-law of $\Gamma=1.63$, absorbed by the Galactic column of
$3.9 \times 10^{20}$~cm$^{-2}$.  The model is a simultaneous fit to
all four instruments separately, while the figure shows the summed SIS
and GIS data for clarity.}
\end{figure}

\begin{table*}[h]
\begin{center}
TABLE 4

Power-Law Fit to the {\sl ASCA\/} Spectra of 1E~0449.4--1823

\begin{tabular}{cccccc}
\tableline
\tableline
 Energy Range& & $N_{\rm H}^{\rm a}$& & &Flux \\
(keV)& $\Gamma$& ($10^{20}$ cm$^{-2}$)& $A^{\rm b}$
& $\chi^2$ (d.o.f.)& (ergs cm$^{-2}$ s$^{-1}$) \\
\tableline
0.5--10 & $1.63^{+0.12}_{-0.09}$ & $<8.8$ & 2.45 & 168.3 (212) &
$1.6 \times 10^{-12}$\\
\tableline
\end{tabular}
\end{center}

\tablenotetext{a}{Column density intrinsic to 1E~0449.4--1823,
in addition to the Galactic column of 
$3.9~\times~10^{20}$~cm$^{-2}$.}

\tablenotetext{b}{Power-law~normalization~at~1~keV~in~the~observed~frame,~in~units~of $10^{-4}$~photons~cm$^{-2}$~s$^{-1}$~keV$^{-1}$ (average of the four instruments).}

\end{table*}

\begin{figure}[h]
\plotfiddle{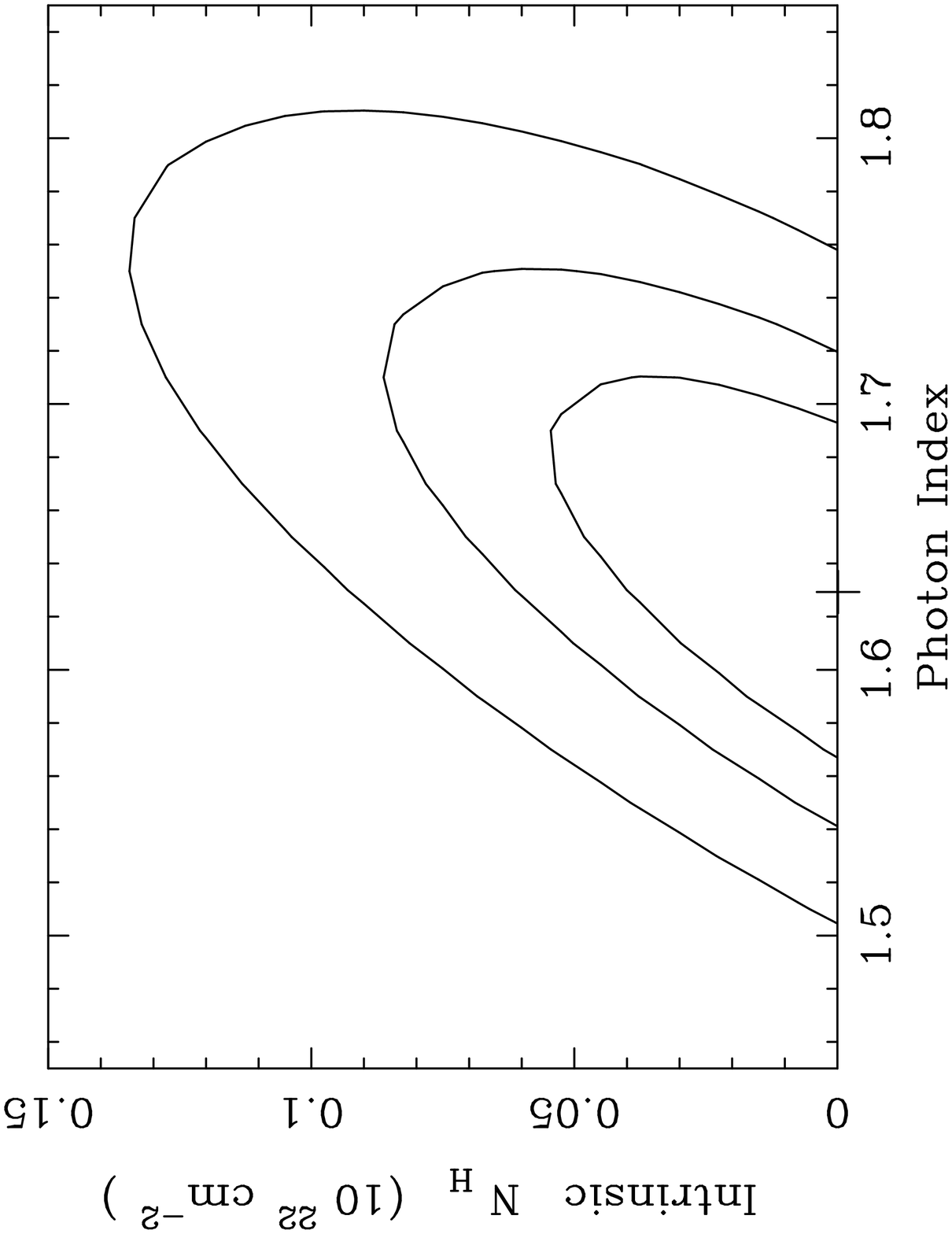}{2.3in}{270}{40}{40}{-150}{220}
\caption{Confidence contours for the parameters of the
power-law model fitted simultaneously to the SIS and GIS spectra of
1E~0449.4--1823.  Contours represent the 68\%, 90\%, and 99\%
confidence limits for two interesting parameters.  Only an upper limit
to the intrinsic column density can be measured, while treating the
Galactic column of $3.9 \times 10^{20}$~cm$^{-2}$ as a fixed
parameter.}
\end{figure}

No emission or absorption features are evident in the residuals from
the power-law fit; the upper limit to the equivalent width of a narrow
Fe~K$\alpha$ emission line at rest frame energy 6.4~keV is $< 440$
eV. The model fit is not improved significantly by the addition of
such a line ($\Delta \chi^{2}$ = 0.01).  In this respect,
1E~0449.4--1823\ differs from the Seyfert~2 galaxies like NGC~1068,
NGC~4945, and NGC~6552 that have fluorescent Fe~K$\alpha$ lines of $EW
= 1.0-1.5$~keV (Marshall et al 1993; Iwasawa et al. 1993; Reynolds et
al. 1994).  Instead, it is more similar to Seyfert~1 galaxies or QSOs.

The {\it ASCA\/}\ X-ray spectrum of 1E~0449.4--1823\ is entirely
consistent with the luminosity measured by {\it Einstein} in the
0.3--3.5 keV band, $5.6 \times 10^{44}$ ergs~s$^{-1}$ (Maccacaro et
al. 1991), and the UV continuum brightness has not increased either
since the observation of Stocke et al. (1983).  Therefore, we have not
even indirect evidence that the growth of the broad emission lines
that we observe in the optical spectrum was caused by an increase in
the intrinsic ionizing luminosity of the nucleus.  However, the
nonsimultaneity of the X-ray and optical observations, and the lack of
regular monitoring during the 15 years since the {\it Einstein}
discovery, make this inference about the absence of causation an
unreliable one.  Nevertheless, we offer an alternative explanation for
the emergence of the broad emission lines in \S 4.

We calculate the X-ray to optical slope $\alpha_{ox}$, defined as
$-{\rm log}\,(f_x/f_o)\,/\,{\rm log}\,(\nu_x/\nu_o)$, where the flux
densities $f_x$ and $f_o$ are calculated at frequencies $\nu_x$ and
$\nu_o$ corresponding to 2~keV and 2500 \AA , respectively, in the
rest frame.  The result is $\alpha_{ox} = 0.93$, smaller than the
value of 1.15 quoted by Stocke et al. (1982), perhaps as a result of
our more accurate measurement of the continuum at 2500 \AA . This
value is unusually small for radio-quiet quasars, implying either a
deficit of UV emission or an X-ray excess.  Typical values are in the
range 1.1--1.7 (Stocke et al. 1990; Boroson \& Green 1992),
indicating, for example, that 1E~0449.4--1823\ could be underluminous
in the UV by a factor of 4 or more.

\eject
\centerline {\bf 4. Discussion}

\centerline {\bf 4.1. So, what is 1E~0449.4--1823 anyway?}

With the discovery of a normal complement of broad emission lines,
there is no longer much to distinguish 1E~0449.4--1823\ from an
ordinary, low-luminosity QSO.  Its absolute magnitude $M_V = -23.5$
meets the QSO criteria, and the equivalent widths of its broad lines
as listed in Table 2 are also in the normal range.  For example, its
broad Mg II equivalent width of 47 \AA\ is comparable to the average
value of 67 \AA\ found by Francis, Hooper, \& Impey (1993), 64 \AA\
found by Zheng et al. (1997), or 34 \AA\ found by Steidel and Sargent
(1991), for radio-quiet quasars in general, and is similar to many of
the individual objects in Corbin \& Boroson (1996).  The X-ray
luminosity of 1E~0449.4--1823\ in the intrinsic 2--10~keV band is $6.7
\times 10^{44}$ ergs~s$^{-1}$, which is also typical of low-luminosity
QSOs.  The upper limit of $9 \times 10^{20}$~cm$^{-2}$ on any
instrinsic column density obscuring the X-ray spectrum allows for
little equivalent visual extinction, $E(B-V) < 0.18$.  Futhermore,
there is nothing unusual about the {\it ratio} of X-ray (2--10 keV) to
broad H$\alpha$ flux, which is $\approx 30$, very close to the mean
value of 40 found for a large sample of Seyfert 1 galaxies by Elvis,
Soltan, \& Keel (1984).  Similarly, the ratio of X-ray to [O~III]
luminosity of 1E~0449.4--1823\ is similar to that of Seyfert~1
and Seyfert~2 galaxies as shown in Figure~5, adapted from Mulchaey
et al. (1994).  However, the large equivalent widths of its
{\it narrow} emission lines are somewhat unusual.  In Figure~6, we
compare several properties of 1E~0449.4--1823\ with those of the 87
low-redshift quasars in the PG sample (Boroson \& Green 1992).
1E~0449.4--1823\ stands out in its [O~III] equivalent width, which is
larger than that of all the PG quasars.  Its FWHM of H$\beta$ is also
at the very high end of the distribution, though it would not be
unusual among radio-loud quasars (Eracleous \& Halpern 1994).

\begin{figure}[h]
\plotfiddle{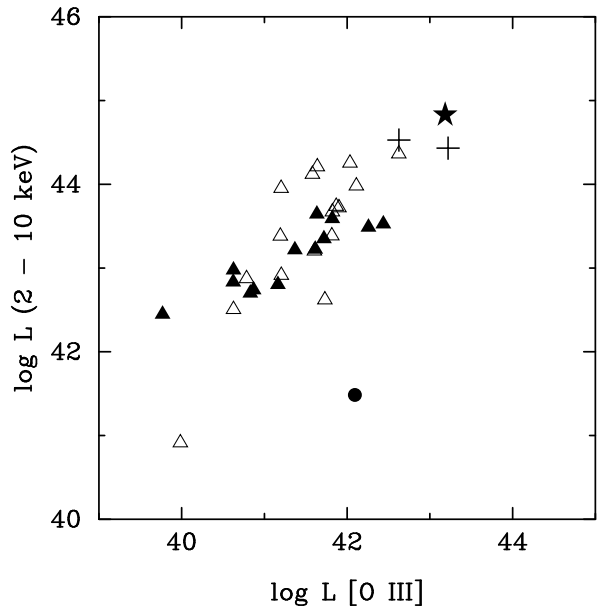}{3in}{270}{45}{45}{-170}{250}
\caption{Properties of Seyfert galaxies and type~2 QSO
candidates with hard X-ray spectra.  Seyfert~1s (open triangles)
and Seyfert~2s (filled triangles) are from Figure~3c of Mulchaey et
al. (1994).  The filled circle is NGC~1068. Crosses are the
ultraluminous {\it IRAS\/}\ galaxies 23060+0505
and 20460+1925
from Brandt et al. (1997) and Ogasaka et al. (1997), respectively.
The star is 1E~0449.4--1823.}
\end{figure}

\begin{figure}[h]
\plotfiddle{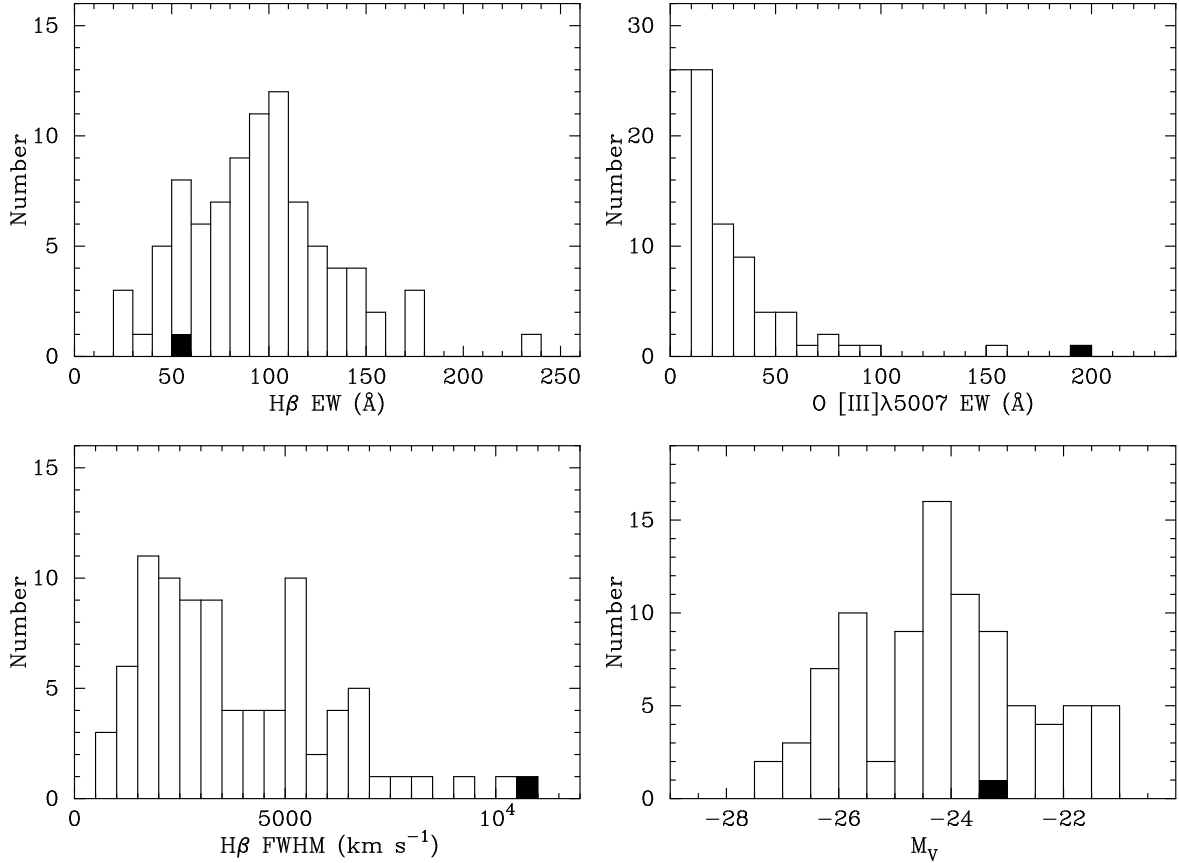}{4.2in}{270}{60}{60}{-230}{350}
\caption{Comparison of the properies of 1E~0449.4--1823\
(dark square) with those of the 87 low-redshift PG quasars from
Boroson \& Green (1992).}
\end{figure}

The original optical spectra from Stocke et al. (1982,1983) and
Stephens (1989) appeared redder than ours, with no evidence for a blue
bump, and broad lines that were weak at best.  The long-term
variability of the Balmer lines and Balmer continuum is obvious,
although difficult to quantify because we do not have the historical
spectra in digital form.  In this respect, the behavior of
1E~0449.4--1823\ resembles that of many Seyfert 1.8 and 1.9 galaxies,
in which the broad Balmer-line components are highly variable on time
scales of years.  Goodrich (1989, 1995) has attributed this effect in
at least some objects to partial obscuration in and around the
broad-line region, which can vary on the dynamical time scale of the
clouds containing the dust.  The variability can be quite dramatic;
some objects originally classified as intermediate Seyferts because of
their barely detectable broad-line components appear years later as
ordinary Seyfert 1 galaxies.  Indeed, if most intermediate type
Seyferts spend only a fraction of their time in a ``low state,'' then
any deliberate survey for them will amass objects that will later
change their classification to Seyfert~1.  And if the probability of
such variable obscuration declines with increasing luminosity, then
this can explain why the rare ``type 2 QSO'' discovered among hundreds
of X-ray sources is likely to revert eventually to an ordinary type~1
spectrum.

We propose, therefore, that 1E~0449.4--1823\ is simply a higher
luminosity example of these variable intermediate Seyfert galaxies. It
is possible that part of the broad-line region of 1E~0449.4--1823\ is
still covered, which could account for the unusual shape of its
emision-line profiles.  Furthermore, if the continuum emitting region
is still partly obscured, which would be consistent with the fact that
the $V$ magnitude estimated from our spectrum has not changed
significantly from the value of $V=18.5$ measured by Stocke et
al. (1992), then this partial obscuration might explain why the
[O~III] equivalent width is so large.  It is likely that the [O~III]
luminosity represents the time-averaged photoionizing flux seen by the
narrow-line region.  If our present line of sight to the continuum is
more obscured than the average one, while the narrow-line region is
unobscured, then both the large equivalent width of [O~III] and the
relatively flat $\alpha_{ox} = 0.93$ could be accounted for by a
depression of the observed UV/optical continuum by about a factor of
4.  Turning now to the X-rays, since the observed X-ray luminosity has
not changed significantly from the {\it Einstein} value, and in the
absence of any evidence for partial covering or reflection in the
X-ray spectrum, it is probably the case that our line of sight to the
X-ray source is unimpeded, and that the observed X-ray luminosity is a
fair representation of its intrinsic value.

An important test of these conclusions can be provided by
spectropolarimetry.  Most intermediate Seyfert galaxies are weakly
polarized, but when they do show polarization it is often variable,
with the continuum and emission lines having different position angles
and wavelength dependence (Goodrich 1989,1995; Martel 1997).  Such
complex behavior, or a lack of polarization altogether, would be
evidence for our hypothesis that we are getting a direct view of at
least parts of the broad-line and continuum emitting regions.  The
alternative view of 1E~0449.4--1823\ as type~2 QSO in unified schemes
would predict uniform polarization across the broad emission lines and
continuum, caused by electron scattering of the light from an
otherwise hidden AGN, with possible dilution from a second,
unpolarized continuum source.  In that case, the broad-emission lines
should be stronger in polarized light than in the direct flux
spectrum, in exact analogy with the hidden Seyfert~1 galaxies (Miller
\& Goodrich 1990, Tran 1995). If it is a hidden QSO, 1E~0449.4--1823\
would then be similar to the hidden Seyfert~1 galaxy Wasilewsi 49
(Moran et al. 1992), which is the only one of its class in which broad
emission-line components are clearly visible in its direct flux
spectrum.  While we believe that the observed change in the broad
emission lines of 1E~0449.4--1823\ already argues against the hidden
QSO model (e.g., the lines have {\it not} been seen to vary in Was
49), the value of independent confirmation via spectropolarimetry is
evident.

\bigskip
\centerline {\bf 4.2. Are there any Type~2 QSOs?}

Having stricken 1E~0449.4--1823\ from the short list of candidates
that are occasionally nominated for the honor of type~2 QSO, it
remains for us to ask if there are {\it any} such objects.  That is,
are there any high-luminosity counterparts of Seyfert~2 galaxies.
Forster \& Halpern (1996) and Halpern \& Moran (1998) recently
addressed this question with regard to the handful of such X-ray
selected objects, offering a generally pessimistic evaluation of the
qualifications of the proposed candidates, and eliminating all but one
from active consideration.  The reader is referred to those papers for
detailed case histories.  The only remaining X-ray selected candidate
that is still claimed to be a type~2 QSO is a very faint emission-line
object, the {\it ASCA\/}\ source AX J08494+4454 at $z=0.9$ (Ohta et
al.\ 1996), In view of its faintness, we consider this object to be no
stronger a candidate than the others which have since been rejected.
At time of this writing, there is a dearth of evidence for the
existence of {\it any} type~2 QSO among X-ray selected AGNs.

Of course, in the standard unified scheme there are no {\it true}
Seyfert~2 galaxies, only Seyfert~1s hidden by molecular tori.  We
repeat here the discussion of Halpern \& Moran (1998) with regard to
the possible implications.  In the unified scheme, the absence of
type~2 QSOs among X-ray selected samples is natural if either 1) the
X-rays from all such objects are hidden from view, or 2) all
sufficiently luminous QSO nuclei are able to remove any obscuring dust
from their vicinity, allowing their broad-line regions to be visible
from any direction.  Although a number of ultraluminous {\it IRAS\/}\
galaxies have Seyfert~2 spectra and hidden broad-line regions (e.g.,
Wills \& Hines 1997, and references therein), there is not much
evidence that they harbor {\it luminous} X-ray sources
(e.g., Brandt et al. 1997; Ogasaka et al. 1997).
The most luminous X-ray source in the nucleus of a
ultraluminous {\it IRAS\/}\ galaxy is {\it IRAS\/}\ 23060+0505, but
its 2--10~keV luminosity is only $1.5 \times 10^{44}$ ergs~s$^{-1}$
(Brandt et al. 1997), well within the range of Seyfert~1 galaxies
as illustrated in Figure~5.
While both of the explanations offered above may be responsible to
some degree for the dearth of type~2 QSOs, there are counterexamples
to either.  First, a substantial number of Seyfert~2 galaxies {\it
are} detected in hard X-rays because their column densities, in the
range $10^{23}-10^{24}$~cm$^{-2}$, are not so large as to be
completely opaque (e.g., Awaki et al.\ 1991; Salvati et al.\ 1997).
Second, even objects of modest quasar-like luminosity, principally
radio galaxies like Cygnus~A, are able to retain their obscuring
material while permitting their broad Mg~II emission lines to be
visible in (Rayleigh) scattered light (Antonucci, Hurt, \& Kinney
1994).  A power-law nuclear X-ray source with 2--10~keV luminosity of
$\sim~1~\times~10^{45}$~ergs~s$^{-1}$ was detected in Cygnus~A by {\it
Ginga} (Ueno et al.\ 1994).  So it~seems that the obscuring gas and
dust that is essential to the unified scheme is neither so substantial
as to prevent direct X-ray detection or indirect UV scattering, nor so
fragile as to be destroyed in the QSO environment.  Thus, the absence
of type~2 X-ray sources of higher luminosity, and the rarity of type~2
QSOs compared to ordinary QSOs, remain significant facts to be
explained whether in the context of unified models or not.  The two
commonly offered explanations mentioned at the beginning of this
paragraph, while having considerable promise, could both use
additional detailed evaluation. In particular, identifications and
careful spectroscopy of serendipitous sources in deep
and {\it AXAF\/}\ surveys should provide the most sensitive test for
hidden QSOs in hard X-rays.  Moderate to high redshifts would aid in
the detectability of highly absorbed sources by shifting their hard
X-rays to lower energy. If no type~2 QSOs are found in these surveys,
the simplest interpretation may be that they do not exist.

\acknowledgments

This work was supported by a grant from NASA under the {\it ASCA}
Guest Investigator Program (NAG~5-2524).  This paper is contribution
646 of the Columbia Astrophysics Laboratory.

\clearpage

\end{document}